\documentclass[showpacs,prb,twocolumn,superscriptaddress,amsmath,amssymb]{revtex4}
\usepackage{amssymb}
\usepackage{color}
\usepackage{graphicx}
\usepackage{epsfig}
\usepackage[caption=false]{subfig}
\usepackage{dcolumn}
\usepackage{bm}
\usepackage{CJKutf8}

\begin{document}
\begin{CJK*}{UTF8}{bsmi}

\title{Quantum Dynamics of a Josephson Junction-Driven Cavity Mode System in the Presence of Voltage Bias Noise}

\author{Hui Wang (王惠)}\affiliation{Department of Physics and Astronomy, Dartmouth College, Hanover, New Hampshire
03755, USA }

\author{M. P. Blencowe}\affiliation{Department of Physics and Astronomy, Dartmouth College, Hanover, New Hampshire
03755, USA }

\author{A. D. Armour}
\affiliation{School of Physics and Astronomy and Centre for the Mathematics and Theoretical Physics of Quantum Non-Equilibrium Systems, University of Nottingham, Nottingham NG7 2RD, UK}

\author{A. J. Rimberg}\affiliation{Department of Physics and Astronomy, Dartmouth College, Hanover, New Hampshire
03755, USA }

\date{\today}

\begin{abstract}
We give a semiclassical analysis of the average photon number as well as photon number variance (Fano factor $F$) for a Josephson-junction (JJ) embedded microwave cavity system, where the JJ is subject to a fluctuating (i.e. noisy) bias voltage with finite dc average.  Through the ac Josephson effect, the dc voltage bias drives the effectively nonlinear microwave cavity mode into an amplitude squeezed state ($F<1$), as has been established previously [A. D. Armour {\it et al}., Phys. Rev. Lett. {\bf 111}, 247001 (2013)], but bias noise acts to degrade this squeezing. We find that the sensitivity of the Fano factor to bias voltage noise depends qualitatively on which stable fixed point regime the system is in for the corresponding classical nonlinear steady state dynamics. Furthermore, we show that the impact of voltage bias noise is most significant when the cavity is excited to states with large average photon number.
\end{abstract}


\maketitle
\end{CJK*}

\section{\label{sec:introduction}Introduction}
When a Josephson junction (JJ) device is embedded within a superconducting microwave cavity, the energy supplied by a dc voltage bias can be converted into microwave photons.\cite{astafiev07,hofheinz11,chen14,cassidy17,westig17} The resulting radiation is predicted to display non-classical properties over a wide range of conditions,\cite{rodrigues07,marthaler11,leppa13,armour13,gramich13,kubala14,armour15,trif15,leppa16,souquet16,dambach16} some of which have now been demonstrated in experiment.\cite{westig17} However, the state of the microwave cavity in such systems is strongly dependent on its quality factor $Q$: for low $Q$ the photons leak out as fast as they are generated, while for high $Q$ a large non-equilibrium photon population can build up. This leads to the interesting question of whether JJ-cavity devices could be used to autonomously generate stable, macroscopic quantum microwave steady states by exploiting the strong induced nonlinearities in the microwave cavity modes.

An effective way to introduce a dc bias into a microwave cavity without compromising its quality factor was demonstrated recently\cite{chen11} and the same design was used in a high-$Q$ factor microwave cavity-embedded Cooper pair transistor (CPT) device\cite{blencowe12} that exhibited lasing behavior for the cavity mode.\cite{chen14} In the latter experiment, a cavity mode was excited to occupations $\sim 100$ photons at co-tunnelling resonances where the system behaves like an effective single JJ. Elsewhere,\cite{armour13} some of us showed using a simple theoretical model of a single JJ embedded in a high-$Q$ cavity that amplitude squeezing might occur even at very large cavity occupation numbers. However, in that analysis the presence of voltage bias noise was neglected, with only cavity loss and associated noise taken into account. Since it is not possible to entirely eliminate bias noise in real devices, it is important to determine its effect on the quantum dynamics and in particular on the predicted amplitude squeezing of the microwave steady states which are expected to be sensitive to various sources of noise. In the present work, we therefore extend the analysis of Ref.~\onlinecite{armour13} to account for bias noise.

Adopting a model in which the JJ feels a classically fluctuating bias voltage, we find that bias noise enters effectively as so-called multiplicative noise, in particular noise terms that multiply the nonlinear interaction terms of the cavity dynamics. This is to be contrasted with the usual additive noise that accompanies the cavity photon decay rate. Using a quantum Langevin equation approach, we investigate the impact of the voltage bias noise on the cavity steady state in the regime where the average photon number  is large so that a semiclassical approximation can be made. The voltage bias noise reduces amplitude squeezing, but in a way that is rather sensitive to the average photon number of the cavity and the dynamical state of the system, changing abruptly when the underlying classical dynamics of the cavity undergoes a bifurcation. Our analysis establishes the conditions which need to be met in order to stabilize non-classical states in JJ-cavity systems containing large photon numbers. We note that the effects of voltage bias fluctuations were accounted for using a complementary quantum master equation approach in a recent study of a JJ-cavity system,\cite{gramich13} but their impact on the quantum dynamics in the regime where the average photon number is large was not investigated.

The outline of this paper is as follows:
in Sec.~\ref{sec:circuit}, we introduce a lumped element circuit model of the JJ-cavity system and  classical Langevin equations that describe the circuit.  In Sec.~\ref{sec:quantumdynamics} we write down the corresponding quantum Langevin equations and describe the semiclassical approximation method. In Sec.~\ref{sec:solutions}, we present the solutions to the average cavity photon number as well as to the fluctuations in the photon number (i.e., Fano factor) and explore their dependence on the key parameters of the system. We also discuss the significance of our results in the context of recent experimental work. Finally, we give some concluding remarks in Sec.~\ref{conclusionsec}. 

\section{\label{sec:circuit} Circuit Model}
\begin{figure}[tb]
\centering
{\includegraphics[width=0.45\textwidth]{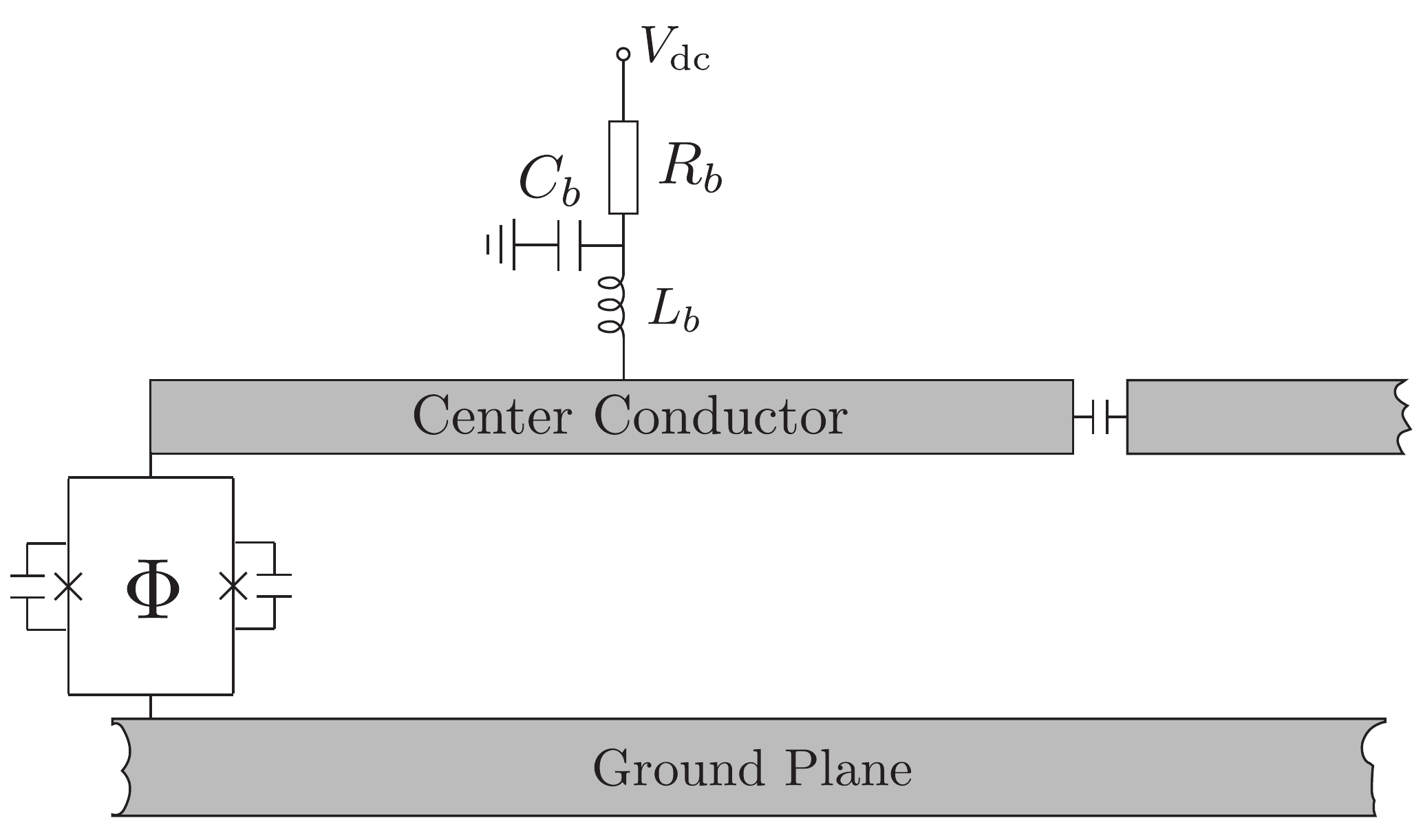}}
\caption{\label{schemefig}Schematic diagram of a $\lambda/2$ microwave cavity-embedded JJ device showing dc bias circuitry (see the main text for details). The center conductor of the cavity is weakly coupled via a capacitor to a probe/transmission line at its right hand end. Note that the upper ground plane of the cavity has been omitted for clarity.}
\end{figure}

The microwave cavity-embedded JJ device\cite{armour13,chen14} is shown schematically in Fig.~\ref{schemefig}. It consists of a parallel combination of two JJs in a SQUID geometry which behaves like an effective  single junction with a flux tunable Josephson critical current $I_J=4I_c\cos\left(\pi\Phi/\Phi_0\right)$, where $I_c$ is the individual JJ critical current (we assume identical JJs for simplicity), $\Phi$ is the external flux bias, and $\Phi_0=h/(2e)$ is the flux quantum.  
The dc voltage bias line connects the cavity center conductor via a series resistance $R_b$ and inductance $L_b$, and a very large parallel bias capacitance $C_b$. Collectively, these lumped circuit elements serve to provide a high impedance, low voltage noise bias line at the several GHz, microwave cavity mode frequency of interest.\cite{chen11}  

\begin{figure}[t]
\centering
{\includegraphics[width=0.45\textwidth]{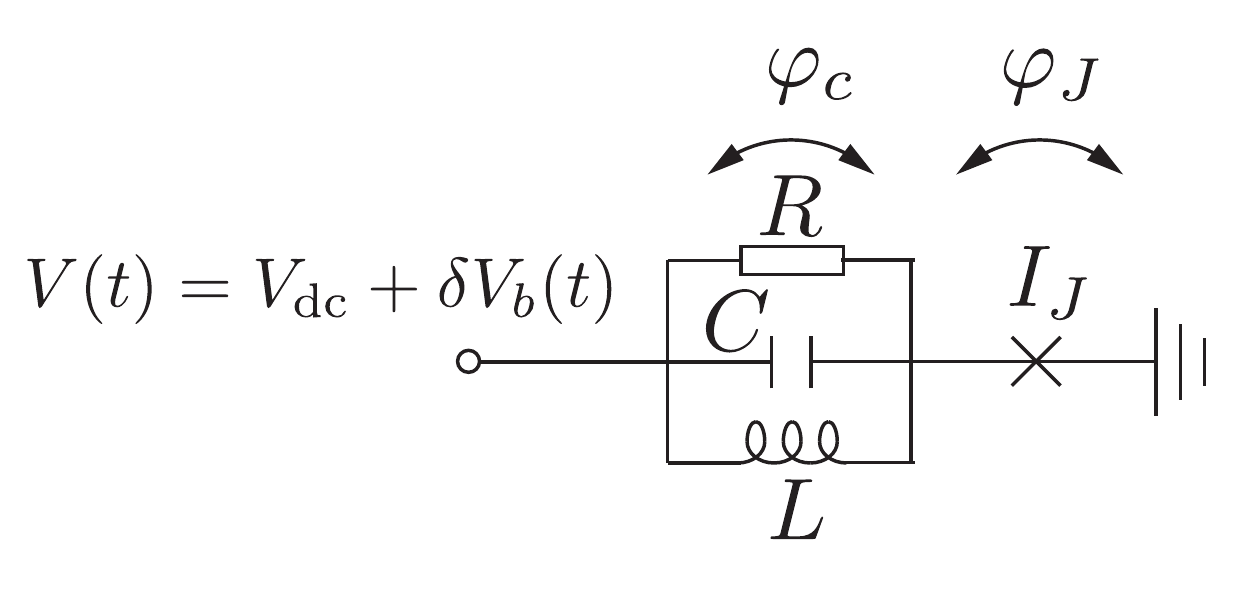}}
\caption{\label{circuitmodelfig}Model lumped element cavity mode-JJ circuit with the various phase coordinate definitions.}
\end{figure}

The model lumped element circuit that we shall in fact analyze is shown in Fig.~\ref{circuitmodelfig}. This circuit captures the essential dynamics of the device scheme in Fig.~\ref{schemefig} within the single mode approximation,\cite{armour13,chen14} where the resistance $R$ takes into account intrinsic cavity damping as well as loss due to capacitive coupling to the probe transmission line shown in Fig.~\ref{schemefig}; the circuit capacitance $C$ takes into account the JJ capacitances as well as the cavity mode effective capacitance.  

Rather than trying to model the specific bias impedance of the device scheme, we shall adopt a more general, phenomenological approach\cite{marthaler11} where we assume a randomly fluctuating, classical voltage bias $V(t)=V_{\mathrm{dc}}+\delta V_b(t)$ (Fig.~\ref{circuitmodelfig}); by definition  $V_{\mathrm{dc}}=\langle V(t)\rangle$ and $\langle \delta V_b(t)\rangle=0$ with angular brackets denoting averaging. Furthermore, we assume for simplicity that the voltage fluctuations are Gaussian distributed with correlation relation
\begin{equation}
\langle \delta V_b (t)\delta V_b (t')\rangle =\delta V_{\mathrm{RMS}}^2 e^{-|t-t'|/
\tau_b},
\label{voltagecorrelation}
\end{equation} 
where $\delta V_{\mathrm{RMS}}=\sqrt{\langle\left( \delta V_b (0)\right)^2\rangle}$ are the root-mean-squared voltage fluctuations and $\tau_b$ gives the decay time of the voltage fluctuation correlations. Such a phenomenological approach allows us to collectively account for the various cryogenic and room temperature voltage noise sources of the actual experimental device that are not captured by the simplified bias circuitry shown in Fig.~\ref{schemefig}.   
      
Applying Kirchhoff's laws to the circuit shown in Fig.~\ref{circuitmodelfig}, we obtain the following equations for the various phase coordinates:
\begin{eqnarray}
\frac{2\pi}{\Phi_0} V(t)&=&\dot{\varphi}_{c}+\dot{\varphi}_J\label{Kirch1eq}\\
C\ddot{\varphi}_c+\frac{1}{R}\dot{\varphi}_c+\frac{1}{L}\varphi_c&=&\frac{2\pi I_J}{\Phi_0}\sin\varphi_J.\label{Kirch2eq}
\end{eqnarray}
Equation~(\ref{Kirch1eq})  allows us to express the JJ phase coordinate $\varphi_{J}$ in terms of the cavity phase coordinate $\varphi_c$ and the bias voltage $V(t)$:
\begin{equation}
C\ddot{\varphi}_c +\frac{1}{L}\varphi_c+\frac{1}{R}\dot{\varphi}_c+\frac{2\pi I_J}{\Phi_0}\sin\left(\varphi_c+\varphi_{b}-\omega_d t\right)=f_c (t),\label{cavityphaseeq}
\end{equation}
where the drive frequency is given by 
\begin{equation}
\omega_d=\frac{2\pi V_{\mathrm{dc}}}{\Phi_0}
\label{drivefreqeq} 
\end{equation}
and the phase bias noise is defined as
\begin{equation}
\varphi_b(t)=-\frac{2\pi}{\Phi_0}\int_0^t dt' \delta V_b(t').
\label{phasebiaseq}
\end{equation}
For completeness, we have also added a cavity `force' noise term $f_c(t)$ which is assumed to be Gaussian distributed and delta function correlated with zero mean:
\begin{equation}
\langle f_c (t) f_c (t')\rangle=\frac{2 k_B T}{R} \left(\frac{2\pi}{\Phi_0}\right)^2 \delta (t-t'),\, \langle f_c(t)\rangle=0,\label{ccorreq}
\end{equation}
where $T$ is the temperature of the cavity environment. This describes the effects of losses from the cavity in the classical description.

The correlation relation for the bias phase noise coordinate $\varphi_b(t)$ follows from the voltage noise correlation relation~(\ref{voltagecorrelation}) (see, e.g., Ref.~\onlinecite{zwanzig}). In the long time limit $t\gg \tau_b$ relevant  for steady state solutions to Eq.~(\ref{cavityphaseeq}), we have
\begin{equation}
\langle \varphi_b(t)\varphi_b(t')\rangle=\gamma_b \left[\Theta (t-t') t' +\Theta (t'-t)t\right],\, \langle\varphi_b(t)\rangle=0,
\label{phasecorrelationeq}
\end{equation}
where the rate constant $\gamma_b$ characterizing the strength of the bias phase noise is
\begin{equation}
\gamma_b =2 \left(\frac{2\pi}{\Phi_0}\right)^2\delta V_{\mathrm{RMS}}^2\tau_b =\frac{1}{2} \left(\frac{2\pi}{\Phi_0}\right)^2 S_{V} (0),
\label{gammabeq}
\end{equation}
with $S_V(0)$ the voltage noise spectral density at zero frequency:
\begin{equation}
S_V(0) =2\int_{-\infty}^{+\infty}d(t-t')\langle \delta V_b (t)\delta V_b (t')\rangle.
\label{spectralVeq}
\end{equation}
Equation~(\ref{cavityphaseeq})  has the form of a nonlinear Langevin equation for the stochastic cavity phase coordinate $\varphi_c(t)$ with noise sources $f_c(t)$ and $\varphi_b(t)$. From Eqs.~(\ref{voltagecorrelation}), (\ref{phasebiaseq}) and (\ref{phasecorrelationeq}), the bias voltage noise and associated phase noise are analogous to the fluctuating velocity and position coordinates respectively of a free Brownian particle; while the mean squared velocity (voltage) is constant in time, the mean squared position (phase) grows linearly in time as the particle randomly wanders throughout its configuration space.   

The classical Hamiltonian for the circuit is
\begin{equation}
H=\left(\frac{2\pi}{\Phi_0}\right)^2\frac{p_c^2}{2C}+\left(\frac{\Phi_0}{2\pi}\right)^2\frac{\varphi_c^2}{2L}
-E_J\cos\left(\varphi_c+\varphi_b -\omega_d t\right)
\label{classicalHeq}
\end{equation}
and the corresponding quantum Hamiltonian can be written as
\begin{equation}
H=\hbar\omega_ca^{\dag}a-E_J\cos\left[\Delta_0(a+a^{\dag})+\varphi_b -\omega_d t\right],
\label{quantumHeq}
\end{equation}
where $a$, $a^{\dag}$ are the annihilation and creation operators respectively for the cavity photons with frequency $\omega_c=1/\sqrt{LC}$, $E_J =I_J\Phi_0/(2\pi)$ is the effective Josephson energy,
and $\Delta_0=(2e^2\sqrt{L/C}/\hbar)^{1/2}$ is the zero-point uncertainty of the cavity mode phase coordinate $\varphi_c$.

For narrow detuning about the bare cavity resonance frequency, $\delta\omega=\omega_c-\omega_d\ll \omega_c$, it is convenient to transform to the frame rotating with the drive frequency $\omega_d$ and then perform the rotating wave approximation (RWA) which should be valid provided $\Delta_0$ and $E_J/(\hbar\omega_c)$ are not too large.\cite{armour13} The unitary operator that transforms to the rotating frame is defined as $U_{\mathrm{RF}}(t)=e^{i\omega_d a^{\dag}a t} U_H(t)$, where $U_H(t)$ is the unitary evolution operator for the full Hamiltionian~(\ref{quantumHeq}). Performing the RWA by dropping rapidly oscillating terms, we have
\begin{equation}
\frac{dU_{\mathrm{RF}}(t)}{dt}\approx -\frac{i}{\hbar} H_{\mathrm{RWA}} U_{\mathrm{RF}}(t),
\label{rfeq}
\end{equation}
where the RWA effective Hamiltonian is
\begin{eqnarray}
&&H_{\mathrm{RWA}}=\hbar\delta\omega\, a^{\dag}a-\frac{i}{2}\tilde{E}_J\cr
&&\times:\left(e^{i\varphi_b}a^{\dag}-e^{-i\varphi_b }a\right)\frac{J_1\left(2\Delta_0\sqrt{a^{\dag}a}\right)}{\sqrt{a^{\dag}a}}:,
\label{HRWAeq}
\end{eqnarray}
with the renormalized Josephson energy $\tilde{E}_J=E_Je^{-\Delta_0^2/2}$, and where $J_1(z)$ is the Bessel function of the first kind and the colons denote normal ordering. Note that the voltage bias noise enters through the random phase factors $e^{\pm i\varphi_b}$ that multiply the anharmonic terms of the Hamiltonian~(\ref{HRWAeq}). From Eq.~(\ref{phasecorrelationeq}) and the assumed Gaussian distributed nature of the fluctuations, it follows that these random phase factors obey the following statistical average relations that will be useful later:
\begin{eqnarray}
\langle e^{\pm i\varphi_b (t)}\rangle &=& e^{-\gamma_b t/2}\cr
\langle e^{\pm i(\varphi_b (t)-\varphi_b (t'))}\rangle&=&e^{-\gamma_b |t-t'|/2}.
\label{phasenoiseaverageeq}
\end{eqnarray}

\section{\label{sec:quantumdynamics}Circuit Quantum Equations}
\subsection{\label{sec:quantumlangevin}Quantum Langevin Equation}
In order to take into account cavity mode damping and associated noise in the quantum dynamics, it is convenient to use the `input-output' approach,\cite{gardiner} 
which assumes weak system-bath couplings in order to justify making a further RWA for the system-bath interaction dynamics as well as a Markov approximation for the bath
dynamics. The rotating frame time dependence of the cavity mode annihilation operator is related to the Heisenberg picture as follows: $a_{\mathrm{RF}}(t)=e^{i\omega_d t}a_H (t)$. The resulting quantum Langevin equation for $a_{\mathrm{RF}}(t)$ is 
\begin{eqnarray}\label{heisenberg}
\frac{da}{dt}&=&-\left(i\delta\omega+\frac{\gamma_c}{2}\right)a-\frac{\Delta_0 {E_J}}{2\hbar}:\bigg[e^{i\varphi_b}J_0\left(2\Delta_0\sqrt{a^{\dag}a}\right)\cr
&&+e^{-i\varphi_b}\frac{a}{a^{\dag}}J_2\left(2\Delta_0\sqrt{a^{\dag}a}\right)\bigg]:+\sqrt{\gamma_c}{a}_{\mathrm{in}},
\label{quantumlangevineq}
\end{eqnarray}
where we have dropped the `RF' subscripts on the mode operators as well as the tilde on $E_J$ for notational convenience, and where $\gamma_c=1/(RC)$ is the cavity mode energy damping rate that accounts for intrinsic mode decay as well as loss due to capacitive coupling to the probe transmission line. The noise operator $a_{\mathrm{in}}(t)$ satisfies the correlation relations $\langle a_{\mathrm{in}}(t)\rangle=0$, $\langle a_{\mathrm{in}}(t) a_{\mathrm{in}}(t')\rangle=0$, $\langle a^{\dag}_{\mathrm{in}}(t) a_{\mathrm{in}}(t')\rangle=0$, and  $\langle a_{\mathrm{in}}(t)a_{\mathrm{in}}^{\dag}(t^{\prime})\rangle=\delta(t-t^{\prime})$, where we approximate the cavity mode environment temperature to be zero (i.e., $k_B T\ll \hbar\omega_c$).    

\subsection{\label{sec:semiclassical}Semiclassical Approach}
A convenient analytical approach to solving Eq.~(\ref{quantumlangevineq}) is the so-called `semiclassical' approximation that involves substituting in $a=\alpha+\delta{a}$, with $\alpha=\langle{a}\rangle$, and then expanding to first order in the  fluctuations\,\cite{armour13} $\delta{a}$ which now include both quantum and classical contributions. The validity of this semiclassical approximation requires that $|\alpha|\gg 1$.  However, because of the classical voltage bias noise, the phase operator $\phi$ in the number-phase expression for the cavity mode operator, ${a}=e^{i{\phi}}\sqrt{{N}}$, becomes completely uncertain with $|\langle e^{i{\phi}}\rangle|\ll 1$; the Wigner function representation of the cavity state would  be circularly symmetric in phase space to a good approximation. Hence $|\alpha|\ll 1$ and so the usual semiclassical approximation method does not apply.

On the other hand, because the classical voltage bias noise is the dominant source of the mode phase operator uncertainty, we expect the phase operator ${\phi}$ and voltage bias phase $\varphi_b$ to be strongly correlated so that $|\langle e^{i({\phi}-\varphi_b)}\rangle|\approx 1$. If we therefore work instead with the transformed mode operator defined as ${b}=e^{-i\varphi_b}{a}=e^{i({\phi}-\varphi_b)}\sqrt{{N}}$, we now have that  $|\langle b\rangle|\gg 1$ when $\sqrt{\langle N\rangle}\gg 1$ and thus the semiclassical approximation can be exploited. In particular, inserting the decomposition $b=\beta +\delta b$ with $\beta =Be^{-i\theta}=\langle b\rangle$ and expanding in $\delta b$ to second order, we have
\begin{eqnarray}
&&\langle e^{i({\phi}-\varphi_b)}\rangle=\langle b N^{-1/2}\rangle\cr
&&=e^{-i\theta}\left[1-\frac{1+2\langle\delta b^{\dag}\delta b\rangle+e^{2i\theta}\langle\delta b^2\rangle-3e^{-2i\theta}\langle\delta b^{\dag  2}\rangle}{8 B^2}\right],\cr
&&\label{phasecorrapprox}
\end{eqnarray}
showing that the validity of the semiclassical approximation in terms of the mode operator $b$ (i.e., $B\gg 1$ and small fluctuations) implies that the phase operator and voltage bias phase are strongly correlated.

From Eqs.~(\ref{drivefreqeq}) and (\ref{phasebiaseq}), the $b(t)$ operator is in fact related to the original Heisenberg picture annihilation operator $a_H(t)$ as follows:
\begin{equation}
b(t)=e^{i\left(\omega_d t-\varphi_b\right)} a_H(t)=\exp\left[i\frac{2\pi}{\Phi_0}\int_0^t dt' V(t')\right] a_H(t).
\label{fullVroteq}
\end{equation}
In particular, $b(t)$ is the mode annihilation operator in the rotating frame defined by the full, fluctuating classical voltage bias $V(t) =V_{\mathrm{dc}}+\delta V_b(t)$.

Substituting the transformed mode operator  ${b}$ into Eq.~(\ref{quantumlangevineq}) and restricting to zero detuning ($\delta\omega=\omega_c-\omega_d=0$)  from now on, we obtain the following quantum Langevin equation for the mode operator ${b}$:
\begin{eqnarray} 
\frac{db}{dt}&=&-\left(\frac{\gamma_c}{2}+i\frac{d\varphi_b}{dt}\right)b-\frac{\Delta_0 {E_J}}{2\hbar}:\bigg[J_0\left(2\Delta_0\sqrt{b^{\dag}b}\right)+\cr
&&\frac{b}{b^{\dag}}J_2\left(2\Delta_0\sqrt{b^{\dag}b}\right)\bigg]:+\sqrt{\gamma_c}{b}_{\mathrm{in}}, 
\label{quantumlangevin2eq}
\end{eqnarray}
where ${b}_{\mathrm{in}}=e^{-i\varphi_b}{a}_{\mathrm{in}}$ satisfies the same correlation relations as ${a}_{\mathrm{in}}$ given above.

The following expectation value identities will turn out to be useful in the analysis below:
\begin{eqnarray}
i\left\langle\frac{d\varphi_b}{dt} b\right\rangle&=&-\left\langle\left(\frac{d}{dt}e^{-i\varphi_b}\right)a\right\rangle=\frac{\gamma_b}{2}\langle b\rangle \label{shapiro1eq} \\
i\left\langle\frac{d\varphi_b}{dt} b^2\right\rangle&=&-\frac{1}{2}\left\langle\left(\frac{d}{dt}e^{-2i\varphi_b}\right)a^2\right\rangle=\gamma_b\left\langle b^2\right\rangle.
\label{shapiro2eq}
\end{eqnarray}
These identities can be derived using the phase factor average relations~(\ref{phasenoiseaverageeq}) and the method of Ref.~\onlinecite{shapiro}. Note that the averaging $\langle\cdots\rangle$ in Eqs.~(\ref{shapiro1eq}) and (\ref{shapiro2eq}) and in the following is with respect to both the classical phase bias noise fluctuations and the quantum fluctuations.

Inserting the decomposition ${b}=Be^{-i\theta}+\delta{b}$ into Eq.~(\ref{quantumlangevin2eq}) and retaining only terms to first order in the fluctuations $\delta b$, we obtain the following equation:
\begin{eqnarray}
\frac{d(Be^{-i\theta})}{dt}+\frac{d\delta{b}}{dt}&=&-\left(\frac{\gamma_c}{2}+i\frac{d\varphi_b}{dt}\right)b+\sqrt{\gamma_c}b_{\mathrm{in}}\cr
&&-\frac{\Delta_0 E_J}{2\hbar}\left[J_0\left(2\Delta_0B\right)+e^{-2i\theta}J_2\left(2\Delta_0B\right)\right]\cr
&&+\frac{i\Delta_0^2E_J}{\hbar}\sin\theta J_1\left(2\Delta_0 B\right)\delta b\cr
&&+\frac{\Delta_0^2E_J}{2\hbar}\left[e^{-i\theta}J_1\left(2\Delta_0B\right)\right.\cr
&&\left.+e^{-3i\theta}J_3\left(2\Delta_0{B}\right)\right]\delta b^{\dag}.  
\label{semiclasslangeq}
\end{eqnarray}

Averaging Eq.~(\ref{semiclasslangeq}), the resulting dynamical equations for the amplitude $B$ and phase $\theta$ are 
\begin{eqnarray}
\frac{dB}{dt}&=&-\frac{\Delta_0E_J}{2\hbar}\cos\theta\left[J_0\left(2\Delta_0 B\right)+J_2\left(2\Delta_0 B\right)\right]\nonumber \\
&&-\frac{1}{2}\left(\gamma_c+\gamma_b\right)B, \label{average1} \\
\frac{d\theta}{dt}&=&\frac{\Delta_0E_J}{2\hbar B}\sin\theta\left[J_0\left(2\Delta_0{B}\right)-J_2\left(2\Delta_0{B}\right)\right]. \label{average2}
\end{eqnarray}
Equations~(\ref{average1}) and (\ref{average2}) differ from those in the absence of phase bias only in the enhanced damping rate $\gamma_c+\gamma_b$.\cite{armour13,meister15}  Evaluating the fixed point solutions to $\dot{B}=\dot{\theta}=0$, for small $E_J$ we have a so-called\cite{armour13} `type-$\mathrm{I}$' stable fixed point which is a solution to
\begin{equation}
B_{\mathrm{I}}=\frac{\Delta_0 E_J}{\hbar\left(\gamma_c+\gamma_b\right)}\left[J_0\left(2\Delta_0B_{\mathrm{I}}\right)+J_2\left(2\Delta_0B_{\mathrm{I}}\right)\right]
\label{typeIeq}
\end{equation}
with $\theta_{\mathrm{I}}=\pi$. Note that the type-$\mathrm{I}$ fixed point depends for its existence on the presence of damping. 

Another type of fixed point--so-called `type-$\mathrm{II}$'--involves the amplitude satisfying
\begin{equation}
J_0\left(2\Delta_0B_{\mathrm{II}}\right)=J_2\left(2\Delta_0 B_{\mathrm{II}}\right),
\label{type2ampleq}
\end{equation} 
with the phase angle determined by 
\begin{equation}
\cos\theta_{\mathrm{II}}=-\frac{\hbar\left(\gamma_c+\gamma_b\right) B_{\mathrm{II}}}{2\Delta_0E_JJ_0\left(2\Delta_0B_{\mathrm{II}}\right)}.
\label{type2phaseeq}
\end{equation}
For this type of fixed point, the average amplitude $B$ is independent of the damping rate and the Josephson energy; from Eq.~(\ref{type2ampleq}), the smallest amplitude type-$\mathrm{II}$ fixed point solution satisfies $2\Delta_0 B_{\mathrm{II}}\approx 1.84$, i.e., $B_{\mathrm{II}}\approx 0.92/\Delta_0$. Increasing $E_J$ from zero, the type-$\mathrm{I}$ fixed point solution eventually becomes unstable at the critical Josephson energy
\begin{equation}
E_{J{\mathrm{crit}}}=E^{(0)}_{J{\mathrm{crit}}}\left(1+\gamma_b/\gamma_c\right),
\label{ejcriticaleq}
\end{equation}
with $E^{(0)}_{J{\mathrm{crit}}}\approx 1.46\hbar\gamma_c/\Delta_0^2$ the value in the absence of bias noise,  undergoing a supercritical pitchfork bifurcation\,\cite{strogatz} 
to two stable type-$\mathrm{II}$ fixed point solutions with equal amplitudes $B_{\mathrm{II}}$ and one unstable type-$\mathrm{I}$ fixed point solution; substituting the critical Josephson energy~(\ref{ejcriticaleq}) into Eq.~(\ref{type2phaseeq}), the critical phase angle is $\theta=\pi$, coinciding with the type-${\mathrm{I}}$ phase angle as it must. As $E_J$ increases beyond its critical value, from Eq.~(\ref{type2phaseeq}) the two bifurcating type-${\mathrm{II}}$ fixed point phase angles approach the values
\begin{eqnarray}
\theta_{\mathrm{II}\,1}&\approx &\frac{\pi}{2}+\frac{E_{J{\mathrm{crit}}}}{E_J},\\
\theta_{\mathrm{II}\,2}&\approx &\frac{3\pi}{2}-\frac{E_{J{\mathrm{crit}}}}{E_J}.
\label{bifurcateangleeq}
\end{eqnarray}

From Eq.~(\ref{semiclasslangeq}), the dynamical equations for the fluctuations about a given fixed point can be concisely written in matrix form as follows: 
\begin{widetext}
\begin{equation}
\left (\begin{array}{c}\dot{\delta b}\\ \dot{\delta b^{\dagger}}\end{array}\right)=\left (\begin{array}{cc}  -i(\nu+\dot{\varphi}_b)-\frac{\gamma_c}{2}& g \\  g^{\ast} &   i(\nu+\dot{\varphi}_b)-\frac{\gamma_c}{2} \end{array}\right)\left (\begin{array}{c}\delta b\\ \delta b^{\dagger}\end{array}\right) +\left (\begin{array}{c}(\frac{\gamma_b}{2}-i\dot{\varphi}_b)Be^{-i\theta}\\ (\frac{\gamma_b}{2}+i\dot{\varphi}_b)Be^{i\theta} \end{array}\right)
+\sqrt{\gamma_c}\left (\begin{array}{c}b_{\mathrm{in}}\\  b_{\mathrm{in}}^{\dagger}\end{array}\right), 
\label{matrix}
\end{equation}
\end{widetext}
where the parameters 
\begin{eqnarray}
\nu&=&-\frac{\Delta_0^2E_J}{\hbar}J_1\left(2\Delta_0 B\right)\sin\theta,\label{nueq} \\
g&=&\frac{\Delta_0^2E_J}{2\hbar}\left[J_1\left(2\Delta_0{B}\right)e^{-i\theta}+J_3\left(2\Delta_0{B}\right)e^{-3i\theta}\right]\label{geq}
\end{eqnarray}
are evaluated at the fixed points.

\section{\label{sec:solutions}Approximate Solutions}
In this section we present analytical results within the semiclassical approximation for the steady state, averaged photon number $\langle N\rangle$ and for the Fano factor, defined as $F=\left(\langle N^2\rangle-\langle N\rangle^2\right)/\langle N\rangle$. The Fano factor provides a measure of the photon number variance; for a coherent state with Poissonian statistics, $F=1$, while for a quantum Fock state, $F=0$. In the absence of voltage bias noise and for realistic device parameters, $F<1$ for both type-${\mathrm{I}}$ and type-${\mathrm{II}}$ fixed point solutions,\cite{armour13} signifying that sub-Poissonian (number squeezed) quantum microwave states are generated in the steady state. Our main concern in what follows will be to determine the extent to which voltage bias noise degrades the squeezing, leading to higher values of $F$, and to explore how this depends on other aspects of the system's dynamics.

\subsection{\label{sec:linearsolns}Linearized Dynamics}
To gain a better understanding of how the semiclassical approximation is applied to calculate $\langle N\rangle$ and $F$, we first consider as a `warm up' exercise the simpler, linearized form of the quantum Langevin equation~(\ref{quantumlangevin2eq}) which is relevant in the small-$E_J$ limit, obtained by setting $J_0\left(2\Delta_0\sqrt{b^{\dag}b}\right)= 1$ and $J_2\left(2\Delta_0 \sqrt{b^{\dag}b}\right)=0$. The resulting equation can be solved exactly:
\begin{eqnarray}
b(t)&=&\int_0^t dt^{\prime}e^{-\frac{\gamma_c}{2}(t-t^{\prime})-i\left[\varphi_b(t)-\varphi_b(t^{\prime})\right]}\cr
 &&\times\left[-\frac{\Delta_0E_J}{2\hbar}+\sqrt{\gamma_c}b_{in}(t^{\prime})\right].
 \label{exactbesolneq}
\end{eqnarray}
The average photon number in the cavity is then obtained as
\begin{equation}
\langle N\rangle=\langle a^{\dag}(t)a(t)\rangle=\langle b^{\dag}(t)b(t)\rangle=\frac{(\Delta_0E_J)^2}{\hbar^2\gamma_c(\gamma_c+\gamma_b)}
\label{lineareq}
\end{equation}
in the steady state limit $t\gg \gamma_c^{-1},\, \tau_b$, where we have used the phase factor average relations~(\ref{phasenoiseaverageeq}) and the $b_{\mathrm{in}}$ correlation relations. 

Alternatively, the averaged photon number can be calculated using the semiclassical decomposition into the linearized form of Eqs.~(\ref{average1}) and (\ref{average2}) for averaged coordinate $\beta$ and Eq.~(\ref{matrix}) for the fluctuation $\delta b$. These linearized equations have the fixed point solutions
\begin{equation}
\beta=-\frac{\Delta_0 E_J}{\hbar(\gamma_c+\gamma_b)}
\label{betasolneq}
\end{equation}
and
\begin{eqnarray}
 &&\delta b(t)=\int_0^tdt^{\prime}e^{-\frac{\gamma_c}{2}(t-t^{\prime})-i\left[\varphi_b(t)-\varphi_b(t^{\prime})\right]}\cr
 &&\times\left[-\frac{\Delta_0 E_J}{\hbar(\gamma_c+\gamma_b)}\left(\frac{\gamma_b}{2}-i\dot{\varphi}_b(t')\right)+\sqrt{\gamma_c}b_{\mathrm{in}}(t^{\prime})\right].
\label{deltabsolneq}
\end{eqnarray}
From Eq.~(\ref{deltabsolneq}), we have in the steady state limit:
\begin{eqnarray}
\langle \delta b^{\dag}(t)\delta b(t)\rangle&=&\frac{\gamma_b}{\gamma_c}\left[\frac{\Delta_0E_J}{\hbar(\gamma_c+\gamma_b)}\right]^2\nonumber\\
&=&\frac{\gamma_b}{\gamma_c}B^2.
\label{bfluctcorreq}
\end{eqnarray}
The semiclassical decomposition expression for the averaged photon number is
\begin{equation}
\langle N\rangle=B^2 +\langle \delta b^{\dag}(t)\delta b(t)\rangle
\label{semiclassavneq}
\end{equation}
and substituting in Eqs.~(\ref{betasolneq}) and (\ref{bfluctcorreq}), we recover the expression~(\ref{lineareq}) for $\langle N\rangle$, which of course we must since the semiclassical decomposition is exact for linear equations.

We now make several observations based on the above analysis of the linearized equations that will be useful for guiding the semiclassical solutions to the nonlinear dynamics in the following section. First, we have that $B\gg 1$ when $\sqrt{\langle N\rangle}\gg 1$, validating the use of the mode operator $b$ for the application of the semiclassical approximation. Second, in contrast to the usual semiclassical approximation in the absence of voltage noise where $\langle N\rangle\approx|\alpha|^2$ with $\alpha=\langle a\rangle$,\cite{armour13} it is essential to include also the fluctuation correlation $\langle\delta b^{\dag}\delta b\rangle$ [see Eq.~(\ref{semiclassavneq})] since, as is apparent from Eq.~(\ref{bfluctcorreq}), the latter can be non-negligible in the presence of classical voltage noise.  Third, comparing Eqs.~(\ref{betasolneq}) and (\ref{bfluctcorreq}), we see that the validity of the semiclassical approximation in expanding to first order in the fluctuations $\delta b$ [Eq.~(\ref{semiclasslangeq})] necessarily requires that the ratio $\kappa=\gamma_b/\gamma_c\ll 1$; in the following section, the parameter $\kappa$ will provide a convenient dimensionless measure of the bias noise strength.

\subsection{\label{sec:nonlinearsolns}Nonlinear Dynamics}
We now return to evaluating the average photon number and Fano factor under conditions where the full nonlinear equations~(\ref{average1}) and (\ref{average2}) for the mean coordinates as well as the full equations~(\ref{matrix}) for the fluctuations must be employed. From Eq.~(\ref{semiclassavneq}), in order to determine the average photon number in the steady state limit, we must solve for the fixed point mean coordinate $\beta=Be^{-i\theta}$ and for the equal time correlation function $\langle \delta b^{\dag}(t) \delta b(t)\rangle$ in the long time limit $t\gg \gamma^{-1}_c,\, \tau_b$ relevant for the steady state behavior. Using Eq.~(\ref{matrix}) and the identities~(\ref{shapiro1eq}), (\ref{shapiro2eq}), an expression for the above correlation function can be obtained by evaluating $d\langle\delta b^{\dag}(t) \delta b(t)\rangle/dt =0$ and $d\langle\delta b(t)^2 \rangle/dt =0$, which give the relations
\begin{eqnarray}
\langle\delta b^{\dag}\delta b\rangle&=& \frac{g^{\ast}}{\gamma_c}\langle\delta b^2\rangle+\frac{g}{\gamma_c}\langle\delta b^{\dag 2}\rangle+\kappa B^2,\label{bdagbeq}\\
\langle \delta b^2\rangle &=&\frac{g/\gamma_c\left(2\langle\delta b^{\dag}\delta b\rangle+1\right)-\kappa B^2 e^{-2i\theta}}{1+2\kappa+2i\nu/\gamma_c},
\label{b2eq}
\end{eqnarray}
where recall $\kappa=\gamma_b/\gamma_c$ and we have used the identity $\langle\delta b\delta b^{\dag}\rangle=\langle\delta b^{\dag}\delta b\rangle+1$. Substituting Eq.~(\ref{b2eq}) and its complex conjugate into Eq.~(\ref{bdagbeq}) and rearranging, we obtain the following expression for the desired correlation function:
\begin{widetext}
\begin{equation}
\left\langle \delta b^{\dag}\delta b\right\rangle=\frac{2|g|^2(1+2\kappa)+\kappa B^2\left\{4\nu^2+\left[\gamma_c(1+2\kappa)\right]^2-e^{-2i\theta}g^{\ast}\left[\gamma_c(1+2\kappa)-2i\nu\right]-e^{2i\theta}g\left[\gamma_c(1+2\kappa)+2i\nu\right]\right\}}{4\nu^2+\left[\gamma_c(1+2\kappa)\right]^2-4|g|^2(1+2\kappa)}.
\label{fluctuation}
\end{equation}
\end{widetext}
In the zero bias noise limit $\kappa=0$, Eqs.~(\ref{b2eq}) and (\ref{fluctuation}) coincide with their corresponding expressions given in Ref.~\onlinecite{armour13}. Note that it is also possible to treat the master equation description of the cavity-JJ system including voltage fluctuations developed in Ref.\ \onlinecite{gramich13} semiclassically and in doing so one recovers entirely equivalent results. 

The combination of $\kappa$ with the classical amplitude term $B^2$ in the numerator of Eq.\ \eqref{fluctuation} means that the voltage bias noise can potentially have a significant impact on the Fano factor in the regime where $B\gg 1$, even if $\kappa$ itself is rather small. This tells us that the impact of voltage bias noise is going to be particularly difficult to control in the `macroscopic' regime where the cavity average photon number is large.

\begin{figure}[thb]
\centering
{\includegraphics[width=0.45\textwidth]{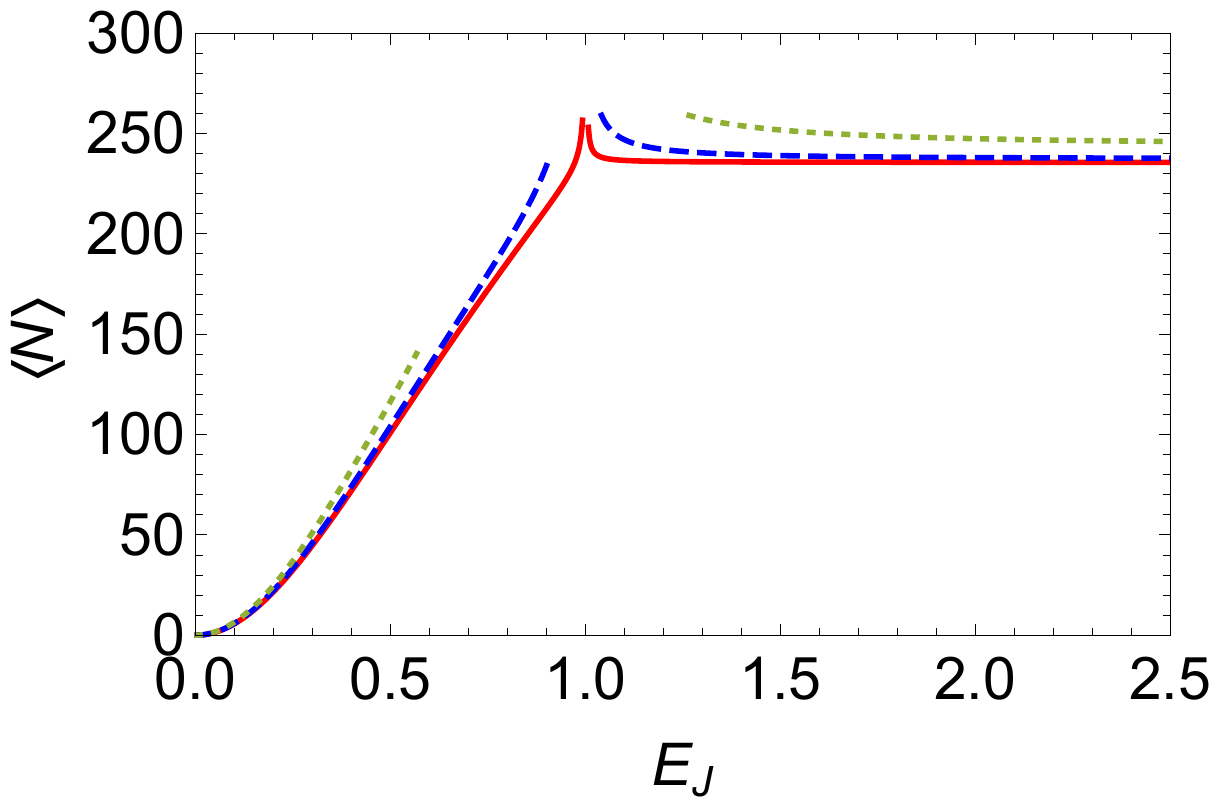}}
\caption{\label{numberfig}Average photon number $\langle N\rangle$ versus $E_J$ (normalized by $E_{J{\mathrm{crit}}}$) for different bias noise strengths $\kappa=0$ (solid line), $0.01$ (dashed line), and $0.05$ (dotted line). The curves are terminated in the neighborhood of the bifurcation points where the semiclassical validity criterion ${\langle \delta b^{\dag}\delta b\rangle}/B^2\leq 0.1$ is violated.} 
\end{figure}

\begin{figure}[thb]
\centering
{\includegraphics[width=0.45\textwidth]{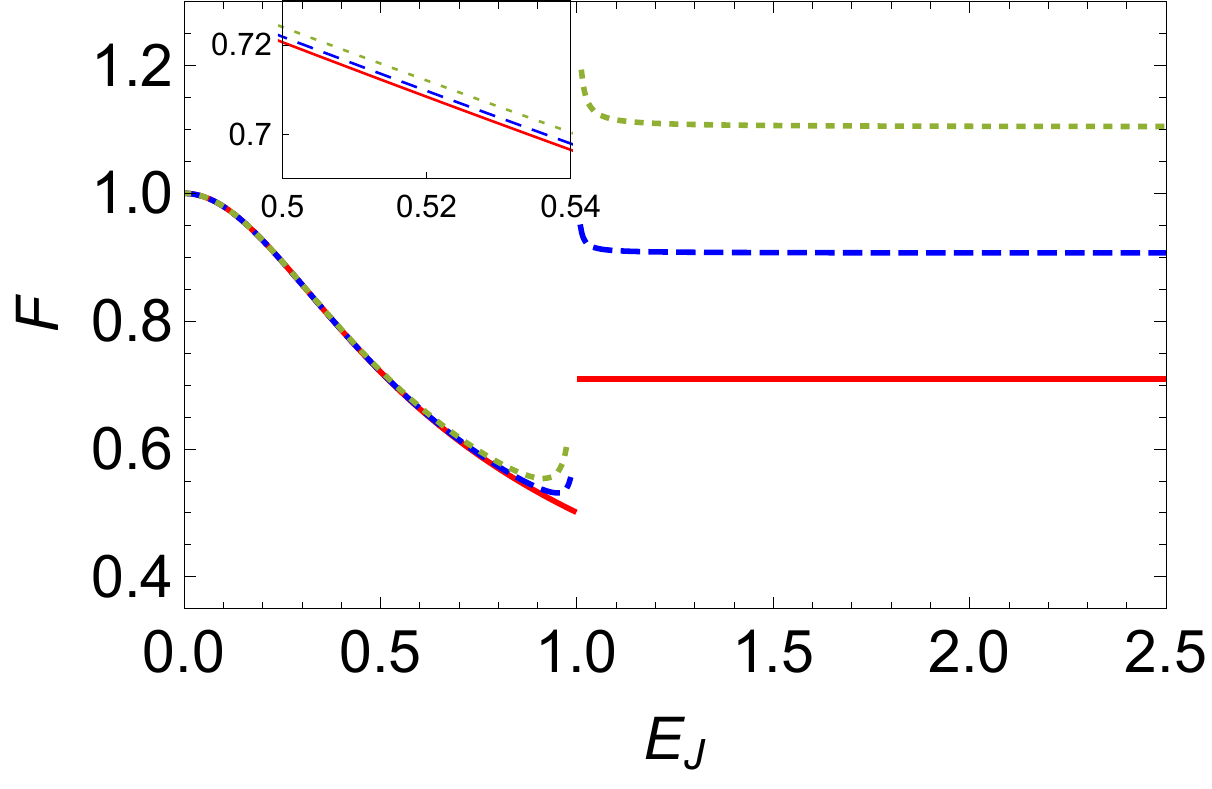}}
\caption{\label{fanofig} Fano factor $F$ versus $E_J$ (normalized by $E_{J{\mathrm{crit}}}$) for different bias noise strengths $\kappa=0$ (solid line), $\kappa=0.001$ (dashed line), and $\kappa=0.002$ (dotted line). The inset shows a zoom-in of the Fano factor versus $E_J$ for a small interval in the type-I stable fixed point regime.}
\end{figure}

Figure~\ref{numberfig} gives the steady state, average photon number $\langle N\rangle$ dependence on the dimensionless Josephson energy ratio $E_J/E_{J{\mathrm{crit}}}$ for a selection of bias noise strengths $\kappa$. The same  parameter values are used as in Ref.~\onlinecite{armour13}, i.e., $\Delta_0=0.06$ and $\omega_c/\gamma_c=10^3$, allowing us to compare our calculations directly in the $\kappa=0$ limit. The critical Josephson energy (\ref{ejcriticaleq}) where the bifurcation from the type-${\mathrm{I}}$ to the type-${\mathrm{II}}$ stable fixed points occurs is then $E_{J{\mathrm{crit}}}/(\hbar\omega_c)\approx 0.40 \,(1+\kappa)$. We adopt as our criterion for the validity of the semiclassical approximation: $\langle \delta b^{\dag}\delta b\rangle/B^2\leq 0.1$; the curves are only indicated for the $E_J$ range where the latter criterion holds. As can be seen in Fig.~\ref{numberfig}, the semiclassical approximation breaks down in the neighborhood of the bifurcation (\ref{ejcriticaleq}) where the type-${\mathrm{I}}$ fixed point becomes unstable.  In particular,  the correlation function $\langle \delta b^{\dag}\delta b\rangle$ grows rapidly near the bifurcation point and the semiclassical breakdown region expands as $\kappa$ increases.   

Moving on to the Fano factor, we have within the semiclassical approximation, where we neglect corrections beyond quadratic order in the fluctuations $\delta b$ and $\delta b^{\dag}$: 
\begin{equation}
F=1+ 2 \left\langle \delta b^{\dag}\delta b\right\rangle+e^{2i\theta}\left\langle\delta b^2\right\rangle+e^{-2i\theta}\left\langle(\delta b^{\dag})^2\right\rangle.
\label{fanofactoreq}
\end{equation}
In Fig.~\ref{fanofig}, we plot the Fano factor as a function of $E_J/E_{J{\mathrm{crit}}}$ for a selection of bias noise strengths $\kappa$. As is to be expected, the Fano factor increases with increasing bias noise. Interestingly, the Fano factor is much less sensitive to bias noise in the type-${\mathrm{I}}$ stable fixed point regime than in the type-${\mathrm{II}}$ stable fixed point regime. 

\begin{figure*}[thp]
\centering
\subfloat[]{\label{fano2figa}
  \includegraphics[width=0.45\textwidth]{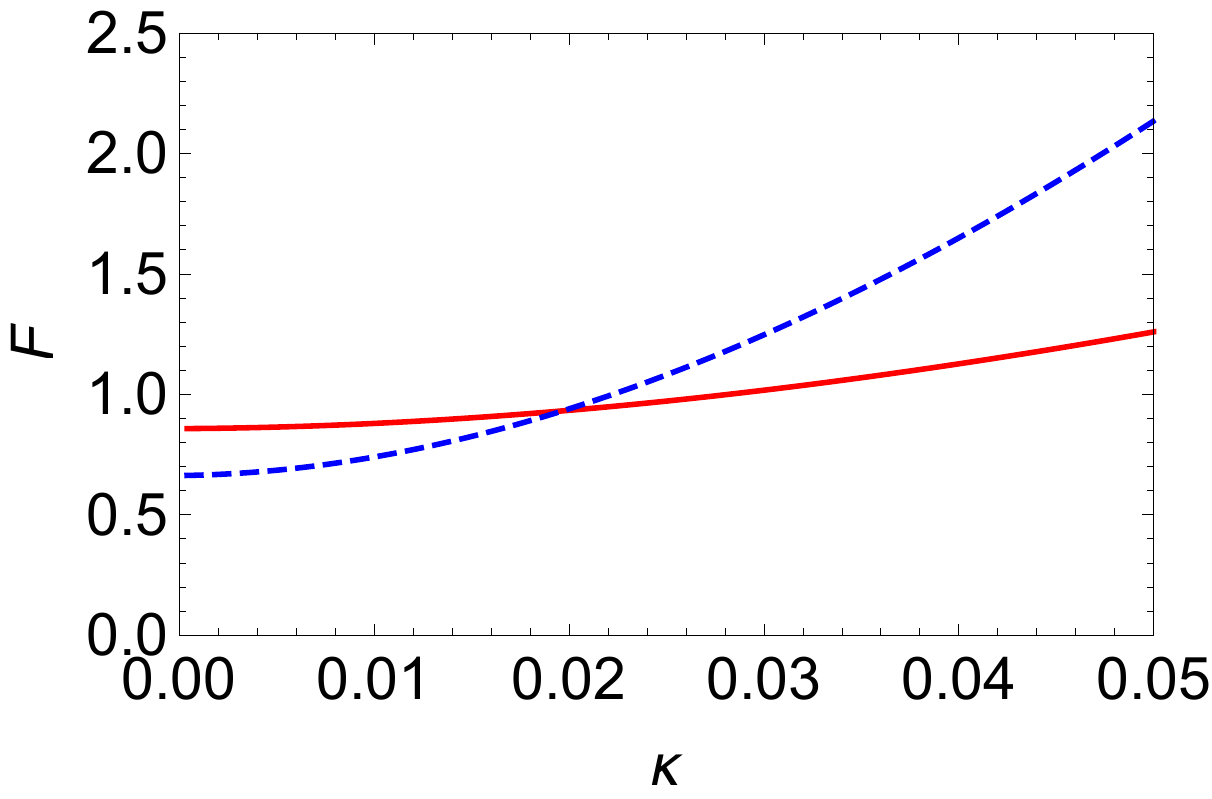}
}
\subfloat[]{\label{fano2figb}
  \includegraphics[width=0.45\textwidth]{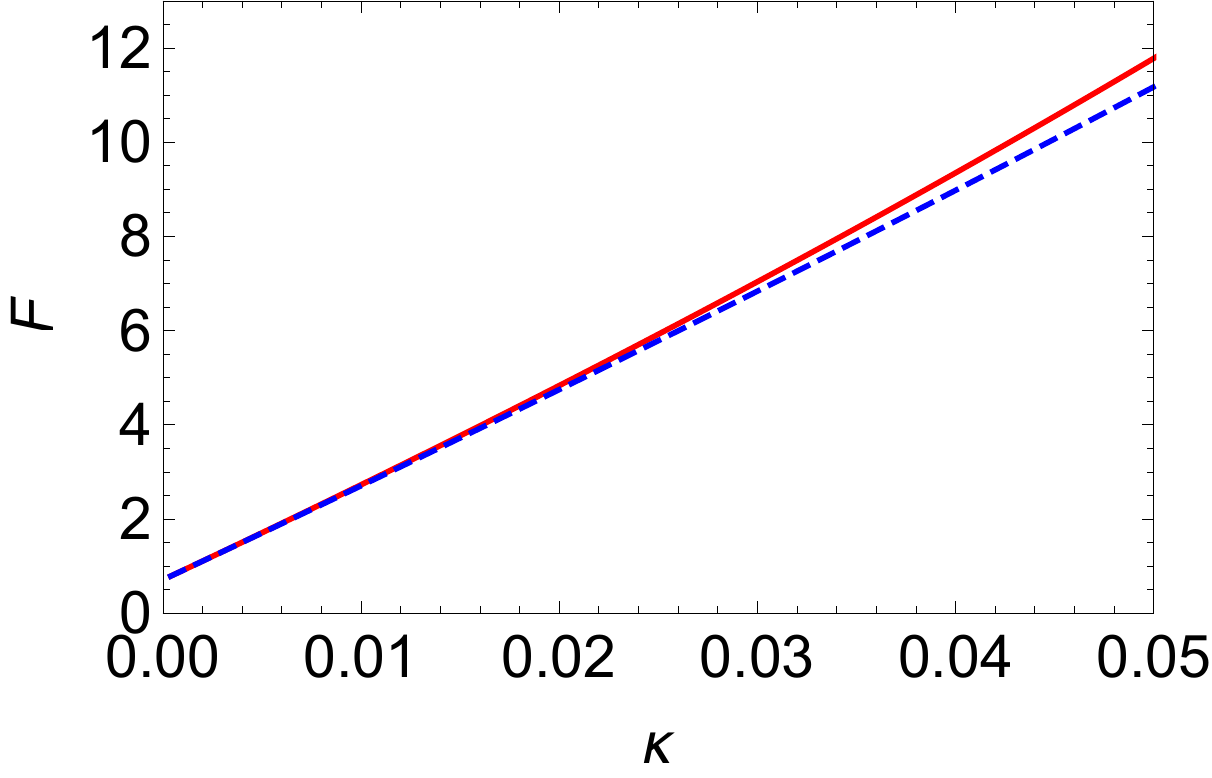}}
\caption{Fano factor $F$ versus bias noise strength $\kappa$ for different $E_J$ (normalized by $E^{(0)}_{J{\mathrm{crit}}}$)  with $\Delta_0=0.06$. (a) Type-$\mathrm{I}$ regime for $E_J=0.30$  (solid line) and $E_J=0.60$ (dashed line).  (b) Type-$\mathrm{II}$ regime $E_J=1.50$ (solid line) and $E_J=2.00$ (dashed line).}
\label{fano2fig}
\end{figure*}

Reviewing the expressions for the Fano factor calculation is instructive for understanding the significant differences between the type-${\mathrm{I}}$ and type-${\mathrm{II}}$ Fano factor sensitivities. From Eqs.~(\ref{b2eq})-(\ref{fanofactoreq}), we expect a dominant $\kappa B^2$ dependence for the Fano factor when $B^2\gg 1$ as is observed in Fig.~\ref{fano2fig}(b). For the type-${\mathrm{I}}$ regime the following simplified algebraic expression for the Fano factor can be derived:
\begin{equation}
F=\frac{1+2\kappa+2g/\gamma_c+4\kappa^2B^2}{1+2\kappa-(2g/\gamma_c)^2}.
\label{fanofactoreq1}
\end{equation}
 Note that, contrary to the general expectation, the leading $\kappa B^2$ dependent term cancels leaving the next-to-leading $(\kappa B)^2$ dependent term when $B^2\gg 1$, which has a relatively weaker, quadratic dependence on $\kappa\ll 1$ as is observed  in Fig.~\ref{fano2fig}(a).

\begin{figure*}[thp]
\centering
\subfloat[]{\label{fano3figa}
  \includegraphics[width=0.45\textwidth]{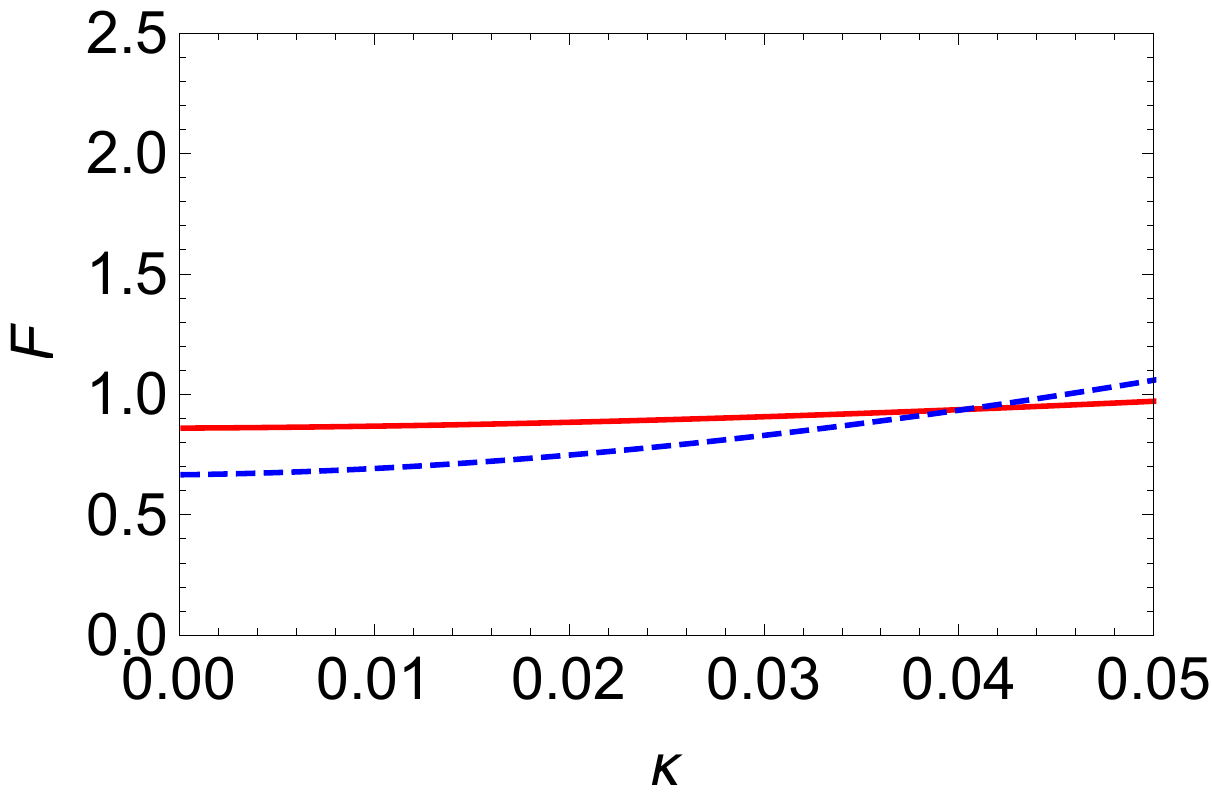}
}
\subfloat[]{\label{fano3figb}
  \includegraphics[width=0.45\textwidth]{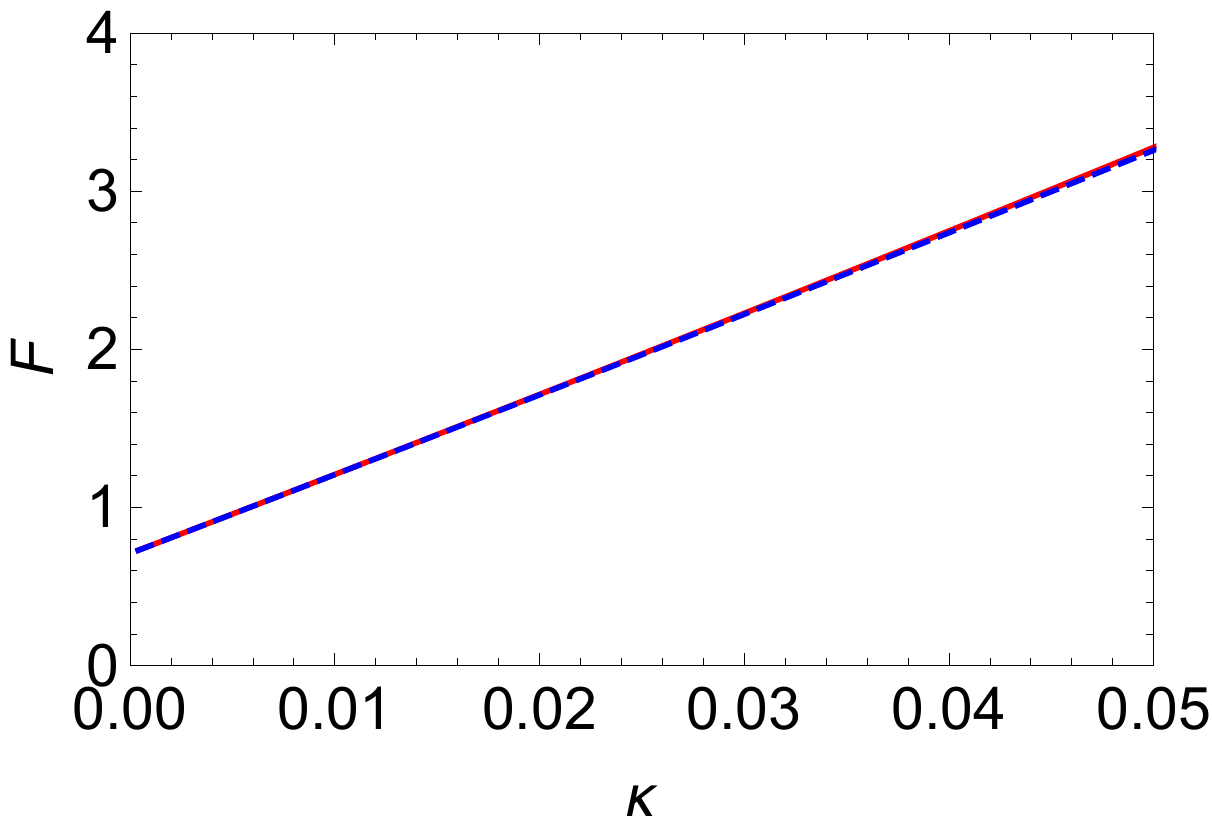}
}
\caption{Fano factor $F$ versus bias noise strength $\kappa$ for different $E_J$ (normalized by $E^{(0)}_{J{\mathrm{crit}}}$)  with $\Delta_0=0.12$. (a) Type-$\mathrm{I}$ regime for $E_J=0.30$  (solid line) and $E_J=0.60$ (dashed line).  (b) Type-$\mathrm{II}$ regime $E_J=1.50$ (solid line) and $E_J=2.00$ (dashed line).}
\label{fano3fig}
\end{figure*}
Figure~\ref{fano3fig} shows the dependence of the Fano factor on $\kappa$ for a larger zero-point uncertainty $\Delta_0=0.12$. 
Increasing $\Delta_0$ reduces the classical amplitude (and hence also the cavity photon number): recall that as $E_J$ increases, the amplitude locks to the value $\simeq 0.92/ \Delta_0$  (see Sec.\ \ref{sec:semiclassical}). Comparing Figs.~\ref{fano2fig} and~\ref{fano3fig}, we see that increasing the value of $\Delta_0$ reduces the rate at which the corresponding Fano factor increases with $\kappa$, thus demonstrating that, as expected,  the effects of voltage bias noise get stronger for larger average photon numbers.

\subsection{Connection with experiment }
As our final consideration, we estimate the Fano factor for the parameters and bias noise values of the experimental device in Ref.~\onlinecite{chen14}. While this device in fact comprised two JJs in series, functioning as a gate-tunable Cooper-pair transistor (CPT), in the operating regime of interest co-tunneling events dominated with only a weak dependence on the CPT gate voltage. Therefore, it is reasonable to model the CPT approximately as a single, effective JJ embedded in the dc voltage biased cavity.  The parameter values for the experimental device were $\Delta_0\approx 0.04$, $\omega_c\approx 2\pi \times 5.3~{\mathrm{GHz}}$, $\gamma_c\approx 9.4 \times 10^6~{\mathrm{s}}^{-1}$, and (neglecting renormalization effects associated with cotunneling\,\cite{Lotkhov}) $E_J/(\hbar\omega_c)\approx 3$, putting the device deep within the type-$\mathrm{II}$ fixed point stability regime with the critical Josephson energy $E_{J{\mathrm{crit}}}/(\hbar\omega_c)\approx 0.26\, (1+\kappa)$ and a predicted Fano factor $F\approx 0.71$ in the absence of bias noise (i.e. $\kappa=\gamma_b/\gamma_c=0$). 

The actual bias noise strength $\kappa$ for the experimental device is a bit more tricky to determine, given that the nature of the voltage bias noise statistics in the experiment is likely more complicated than our simple assumed form for the correlation relation~(\ref{voltagecorrelation}). Furthermore, actual measurements of the cavity mode operator $a(t)$ involve frequency filtering and time domain averaging, so that the various equal time correlation calculations carried out in the present work would need to be replaced by more involved, non-equal time correlation calculations in order to evaluate the Fano factor for the experimental device. Nevertheless, as a rough estimate we can assume that $\gamma_b$ is given by the measured linewidth of the cavity power emission spectrum on resonance: $\gamma_b\approx 2\pi\times 70~{\mathrm{kHz}}$, so that $\kappa\approx 0.05$. With this value for $\kappa$, we find that $F>1$ for both the type-I and type-II stable fixed point regimes; in order to have $F<1$, we require $\kappa<0.01$ in the type-I regime and $\kappa<0.001$ in the type-II regime. Thus, in experiments on simpler single JJ devices with flux tunable Josephson energy,\cite{armour13,cassidy17} the best chance to observe  $\langle N\rangle\gg 1$ microwave photon states with $F<1$ will be in the type-I regime just below the critical Josephson energy, although the voltage bias noise will need to be controlled rather better than in Ref.~\onlinecite{chen14}. In particular, the contribution to the zero frequency spectral noise density $S_V (0)$ [see Eqs.~(\ref{gammabeq}) and (\ref{spectralVeq})] from the cryogenic part of the experimental device may be reduced by decreasing the bias resistance $R_b$ and increasing the the bias capacitance $C_b$. On the other hand, by employing a lower noise, first stage cryogenic microwave amplifier, shorter measurement averaging times are required. Hence there is less jitter from the room temperature voltage source part.   

\section{\label{sec:conclusions}conclusions}
\label{conclusionsec}
We have analyzed the average photon number as well as Fano factor giving the photon number variance for a noisy voltage biased, embedded Josephson junction microwave cavity system. The embedded Josephson junction induces an effective nonlinearity in the cavity mode dynamical equations, as well as a tunable drive tone through the ac Josephson effect. The voltage bias noise was found to have a different order of magnitude effect on the Fano factor, depending on which stable fixed point regime the system is in for its steady state nonlinear dynamics. We also found that the voltage bias noise has more impact when the cavity average photon number is large, making it more difficult to produce number-squeezed states.
However, by quantifying the effects of the voltage bias noise, the present work makes clear the conditions that will need to be met in order for future experiments to demonstrate steady, macroscopic quantum amplitude squeezed microwave states of light.
   
\section*{Acknowledgements}
This work was supported by the NSF under Grants No. DMR-1507383 and DMR-1507400, and by the ARO under Grant No. W911NF-13-1-0377.


\begin{thebibliography}{99}
\bibitem{astafiev07} O. Astafiev, K. Inomata, A. O. Niskanen, T. Yamamoto, Yu. A. Pashkin, Y. Nakamura and J. S. Tsai, Nature (London) {\bf 449} 588 (2007).
\bibitem{hofheinz11} M. Hofheinz, F. Portier, Q. Baudouin, P. Joyez, D. Vion, P. Bertet, P. Roche, and D. Esteve, Phys. Rev. Lett. {\bf 106}, 217005 (2011).
\bibitem{chen14} F. Chen, J. Li, A. D. Armour, E. Brahimi, J. Stettenheim, A. J. Sirois, R. W. Simmonds, M. P. Blencowe,
and A. J. Rimberg, Phys. Rev. B {\bf 90}, 020506 (2014).
\bibitem{cassidy17} M. C. Cassidy, A. Bruno, S. Rubbert, M. Irfan, J. Kammhuber, R. N. Schouten, A. R. Akhmerov and L. P. Kouwenhoven, Science {\bf 355}, 939 (2017).
\bibitem{westig17} M. Westig, B. Kubala, O. Parlavecchio, Y. Mukharsky, C. Altimiras, P Joyez, D. Vion, P. Roche, M. Hofheinz, D Esteve, M Trif, P. Simon, J. Ankerhold and F. Portier, arXiv:1703.05009 (2017). 
\bibitem{rodrigues07}  D. A. Rodrigues, J. Imbers and A. D. Armour, Phys. Rev. Lett. {\bf 98} 067204 (2007).    
\bibitem{marthaler11}M. Marthaler, J. Lepp\"{a}kangas, and J. H. Cole, Phys. Rev. B {\bf 83}, 180505(R) (2011).
\bibitem{leppa13} J. Lepp\"{a}kangas, G. Johansson, M. Marthaler and M. Fogelstr\"{o}m,  Phys. Rev. Lett. {\bf 110}, 267004 (2013).
\bibitem{armour13} A. D. Armour, M. P. Blencowe, E. Brahimi, and A. J. Rimberg, Phys. Rev. Lett.
{\bf 111}, 247001 (2013).
\bibitem{gramich13} V. Gramich, B. Kubala, S. Rohrer and J. Ankerhold, Phys. Rev. Lett. {\bf 111}  247002 (2013).
\bibitem{kubala14}  B. Kubala, V. Gramich and  J. Ankerhold, Phys. Scr. {\bf T165},  014029 (2015).
\bibitem{armour15} A. D. Armour, B. Kubala, and J. Ankerhold, Phys. Rev. B {\bf 91}, 184508 (2015).
\bibitem{trif15} M. Trif and P. Simon, Phys. Rev. B {\bf 92}, 014503 (2015).
\bibitem{leppa16} J. Lepp\"{a}kangas, M. Fogelstr\"{o}m, M. Marthaler and G. Johansson,   Phys. Rev. B {\bf 93}, 014506 (2016).
\bibitem{souquet16} J.-R. Souquet and A. A. Clerk, Phys. Rev. A  {\bf 93}, 060301 (2016).
\bibitem{dambach16} S. Dambach, B. Kubala, and J. Ankerhold, New J. Phys. {\bf 19}, 023027 (2017).
\bibitem{chen11} F. Chen,  A. J. Sirois, R. W. Simmonds, A. J. Rimberg, Appl. Phys. Lett. {\bf 98}, 132509 (2011).
\bibitem{blencowe12}M. P. Blencowe, A. D. Armour, and A. J. Rimberg, in {\it Fluctuating Nonlinear Oscillators: From Nanomechanics to Quantum Superconducting Circuits}, ed. M. Dykman
(Oxford University Press, Oxford, 2012).
\bibitem{zwanzig}R. Zwanzig, {\it Nonequilibrium Statistical Mechanics} (Oxford University Press, Oxford, 2001).
\bibitem{gardiner}C. W. Gardiner and M. J. Collett, Phys Rev A {\bf 31}, 3761 (1985).
\bibitem{shapiro} V. E. Shapiro and V. M. Loginov, Physica A {\bf 91}, 563 (1978).
\bibitem{meister15} S. Meister, M. Mecklenburg, V. Gramich, J. T. Stockburger, J. Ankerhold, and B. Kubala Phys. Rev. B {\bf 92} 174532 (2015).
\bibitem{strogatz}S. H. Strogatz, {\it Nonlinear Dynamics and Chaos}, 2nd ed. (Westview Press, Boulder, 2015).
\bibitem{Lotkhov} S. V. Lotkhov, S. A. Bogoslovsky, A. B. Zorin, and J.
Niemeyer, Phys. Rev. Lett. {\bf 91}, 197002 (2003).
\end{thebibliography}
\end{document}